# Reimagining Voltage-Controlled Cryogenic Boolean Logic Paradigm with Quantum-Enhanced Josephson Junction FETs

Md Mazharul Islam[1], Diego Ferrer[1], Shamiul Alam[1], Juan P. Mendez[3], Denis Mamaluy[3], Wei Pan[2], and Ahmedullah Aziz[1*]

[1] Dept. of Electrical Eng. and Computer Sci., University of Tennessee, Knoxville, TN, 37996, USA
[2] Sandia National Laboratories, Livermore, CA 94550, USA
[3]Sandia National Laboratories, Albuquerque, NM 87123, USA
[*]Corresponding Author. Email: aziz@utk.edu

***Abstract-* The growing demand for ultra-low-power computing and the emergence of quantum technologies have intensified interest in cryogenic electronics, particularly superconducting devices. Despite their promise, current-controlled superconducting components face fundamental challenges in cascadability, limiting their effectiveness in complex logic architectures. To overcome this, recent efforts have focused on developing gate-tunable superconducting devices, such as Josephson junction field-effect transistor (JJFET). However, achieving robust control and sufficient supercurrent gain—both critical for transistor-like performance in logic circuits—remains a key challenge. A recent advancement in JJFET design, based on InAs/GaSb heterostructures, demonstrates enhanced gain and favorable device characteristics suitable for circuit integration. Building on this innovation, we propose and analyze fundamental voltage-controlled logic topologies using the quantum-enhanced JJFET. We develop a Verilog-A-based circuit compatible compact model of the quantum enhanced JJFET which accurately captures the experimentally observed device characteristics. To ensure cascadability, our logic circuits incorporate the nanocryotron (nTron), a superconducting nanowire-based thermal switch. Through simulation-based analysis, we demonstrate the successful implementation of fundamental logic gates, including NOT, NAND, and NOR. Furthermore, we design a 3-input majority gate, which plays a pivotal role in quantum and reversible computing due to its universality. Finally, to demonstrate the cascadability of our proposed logic topology, we demonstrate the operation of a 2-input XOR gate based on our designed JJFET-based NOT, NAND, and NOR gate.**

## Introduction

The end of Moore's Law and the plateauing of traditional silicon-based scaling [1] have driven the exploration of alternative device platforms that can deliver higher energy efficiency and performance beyond conventional CMOS [2]. In this context, cryogenic electronics, particularly superconducting devices, have emerged as promising candidates due to their ability to operate with unprecedented energy dissipation and ultrafast switching speeds [3]–[5]. A recent demonstration of a processor prototype based on superconducting devices has been shown to consume ~80 times lower power compared to the conventional CMOS-based counterpart [6]. In addition to their appeal for classical computing, superconducting devices are inherently compatible with quantum technologies for their cryogenic operating temperature [7], [8], making them well-suited for quantum and hybrid classical-quantum computing systems operating at cryogenic temperatures.

Despite their advantages, most existing superconducting logic circuits rely on current-controlled devices, which suffer from limited fan-out capability and poor voltage gain[9]–[11]. This severely restricts their scalability and integration into large-scale logic systems. To address this issue, significant efforts have focused on developing voltage-controlled superconducting devices that could mimic the functionality of transistors, enabling logic design methodologies analogous to CMOS[12]. Among the most promising candidates are **Josephson junction field-effect transistors (JJFETs)**, which replace the semiconductor channel in a MOSFET with a weak-link superconducting region whose supercurrent can be modulated by a gate voltage [13], [14]. The key operational principle of JJFETs lies in the gate-tunable carrier density $n$, which affects the coherence length $\xi_C$ and, consequently, the Josephson critical current $I_c$However, conventional JJFETs, based on classical semiconducting channels, exhibit a weak dependence of the superconducting coherence length ($\xi_C$) on carrier density ($n$) ($\xi_C \propto n^{\gamma}, \gamma$ =0.5 for classical semiconductor channel material). This results in limited gate control over the critical current ($I_C$) yielding a small gain factor(), $\alpha_R = \frac{dI_C}{dVg} \cdot \frac{\pi\Delta}{I_C}$, (here, $\Delta$ is the superconducting gap) which falls far below the unity threshold required for Boolean logic operation [15]. To address this limitation, a recent study introduced a quantum-enhanced JJFET leveraging a zero-gap InAs/GaSb heterostructure [16]. This material system supports a gate-tunable excitonic insulator (EI) phase transition [17]–[19], which induces a sharp variation in carrier coherence length with gate voltage, yielding a significantly larger gain factor $\alpha_R \sim 0.06$, over 50 times greater than that of InAs-based JJFETs [20]. While this gain factor remains below unity, Ref. [16] outlines several pathways for further enhancement, including (i) reducing the junction length, (ii) replacing Ta with Al to exploit lower gap superconductors, and (iii) optimizing heterostructure growth to sharpen the EI transition. However, even with a gain factor lower than unity, we show that quantum-enhanced JJFETs can be successfully utilized to implement logic gates by leveraging their intrinsically nonlinear gate response and improved modulation depth. The underlying physics in this device relies on the Coulomb-driven condensation of electron-hole pairs in the charge neutrality regime, enabling a highly nonlinear and sensitive modulation of $I_C$.

In this work, we propose and analyze fundamental voltage-controlled superconducting logic circuits based on the quantum-enhanced JJFET. We develop a circuit-compatible Verilog-A-based compact model that accurately captures the experimentally observed device characteristics reported in [16]. To achieve fanout capability in our proposed logic circuits, we employ superconducting three terminal device known as **nanocryotron (nTron)** [21], [22]—a superconducting nanowire-based thermal switch to ensure the input and output voltage matches. Using simulation-based analysis in HSPICE, we demonstrate key logic primitives, including NOT, NAND, NOR, and a 3-input majority gate, which is particularly important for quantum and reversible computing[23], [24]. Our proposed logic topologies potentially solve the issue of single fan-out and the need for extra current splitter circuits in current-controlled superconducting logic designs[25], [26]. To showcase that our proposed voltage-controlled logic topologies can be cascaded in multiple stages, we further analyze a 2-input XOR gate using the implemented JJFET-based gates.

## Device Characteristics and Modeling Approach of Josephson Junction Field Effect Transistor

The quantum-enhanced Josephson junction field-effect transistor (JJFET) investigated in this work is based on a zero-gap InAs/GaSb heterostructure channel contacted by superconducting tantalum (Ta) electrodes. The device leverages an excitonic insulator (EI) phase transition to achieve enhanced gate tunability of the superconducting critical current, $I_C$, essential for implementing voltage-controlled logic. The effective channel length between the superconducting contacts is approximately 500 nm.

A sharp onset of superconducting behavior is observed near $V_G$–$V_T$ =0.24 V (or $V_G$ =- 0.46 V), beyond which, critical current ($I_C$) increases rapidly (Fig. 1(e)). Below this threshold, $I_C$ vanishes, and the device remains resistive—mimicking digital OFF/ON states. The experimentally observed $\frac{dI_C}{dV_G}$ in the transition region is approximately 33 µA/V. Notably, a true zero-voltage supercurrent state is not observed in the *I*–*V* characteristics. Instead, the device exhibits a low-resistance regime at lower bias, transitioning sharply to a higher-resistance state at larger currents. This behavior is attributed to the superconducting proximity effect, where superconducting correlations from the Ta-electrodes induce partial superconductivity in the semiconducting channel. The finite resistance observed even in the low-bias regime is likely due to the long channel length (500 nm), which weakens Josephson coupling and prevents full phase coherence. The two distinct resistance levels observed below and above the critical current $I_C$, are strongly dependent on the gate voltage $V_G$, and are denoted as $R_{SG}$ (sub-gap resistance) and $R_N$ (normal-state resistance), respectively, as illustrated in Fig. 1(d). To enable circuit-level simulations, we developed a Verilog-A-compatible compact model that captures the $V_G$-dependent behavior of both the channel resistances and $I_C$, calibrated against the experimental data illustrated in Figs. 1(d) and 1(e). The resistances $R_{SG}$ $(V_G)$ and $R_N$ $(V_G)$ are modeled phenomenologically using a look-up table-based approach, allowing accurate prediction over the measured voltage range. The gate-dependent critical current is modeled using a piecewise linear fit that reflects the abrupt onset of superconductivity at the critical voltage $V_{GT}$=0.24 V (Fig. 1(e)). The expression for $I_C$ is given by:

$$I_C = \left(\frac{1 + \ sign(V_{GT} - 0.24)}{2}\right) \times (2.886 \times 10^{-7} + 3.21 \times 10^{-7} \times V_{GT})$$

Where, $V_{GT}$ = $V_G$ - $V_T$, is the gate overdrive voltage and $V_T$ = -0.7 V is the channel threshold voltage. The corresponding drain voltage, $V_D$ is modeled based on whether the drain current, $I_D$ is below or above $I_c$:

$$V_D = \begin{cases} I_D \times R_{SG}(f(V_{GT})) & ; I_D < \ I_C \\ I_D \times R_N(f(V_{GT})) & ; I_D \geq I_C \end{cases}$$

A comparison between the modeled and experimental *I*–*V* characteristics is illustrated in Fig. 1(f), demonstrating that the compact model accurately reproduces the nonlinear switching behavior and resistance transitions observed in the device.

**JJFET-based Boolean Logic Circuit Design**

For any logic circuit to be practically useful, its output must be capable of **d**riving subsequent logic gates, a property essential for fanout capability. Here, in the case of the quantum-enhanced JJFET, the generated drain voltage ($V_D$) remains in the millivolt range (Fig. 1(f)), whereas achieving a significant change in drain current ($I_D$) requires altering the gate voltage ($V_G$) by hundreds of millivolts (Fig.1(f)). To address this mismatch and ensure seamless logic-level interfacing between gates, we employ a three-terminal superconducting electrothermal switch known as the **nanocryotron (nTron)** (Fig.1(g)). The nTron operates by injecting a gate current ($I_G$) into a narrow choke region, where localized Joule heating generates resistive hotspot. This hotspot raises the local temperature and suppresses superconductivity in the adjacent channel region, thereby reducing the critical current ($I_C$). When a bias current greater than the suppressed $I_C$ is applied, the channel undergoes a transition to the resistive state (Figs.1(h,i)). Thus, a small $I_G$ can modulate a much larger channel switching current (Fig.1(i)), providing digital gain and enabling fanout. This mechanism allows the nTron to act as a voltage-controlled switch or amplifier in cryogenic logic circuits without requiring Josephson junctions. We have developed a Verilog-A based compact model for the nTron and calibrated it with experimental data reported in [21]. In our JJFET-based logic design, a logic '0' is represented as -0.7V and logic '1' by 0V at the gate terminal. Thus, we require 0.7V separation between logic '0' and '1' at the output of the nTron. Given that the reported channel resistance of the nTron is approximately ~20 kΩ, we choose an nTron channel bias of 35 $\mu A$ to produce the required voltage drop (Fig.1(j)).

Fig.2(a) illustrates the circuit topology of our designed Copy gate. When the input voltage $V_{IN}$ is low (-0.7 V, $V_{GT}$ = 0V), the JJFET exhibits high resistance diverting a large portion of $I_{Bias1}$ through the parallel resistor ($R_P$) (Fig.2(b)). The resulting current through the nTron gate is insufficient to switch the nTron channel and the channel remains superconducting generating 0V across it (Fig. 2(b)). In contrast, when $V_{IN}$ = 0V ($V_{GT}$ = 0.7 V), the JJFET exhibits lower resistance drawing sufficient current from the $I_{Bias1}$ ($I_G$ > 10.3 $\mu$A) (Fig. 2(c)). This drives the nTron channel to switch from superconducting to resistive state (Fig.2(c)). With a channel resistance of ~20 kΩ and a bias current of $I_{Bias2}$=35$\mu$A, the nTron generates an output of 0.7 V (Fig. 2(c)). To maintain consistency in voltage levels between the output and the input of the next logic gate, a bias voltage of $V_{bias}$= -0.7 V is applied at the nTron channel. This biasing approach is consistently used across all our proposed logic topologies to ensure proper fanout and voltage-level compatibility. The series and parallel resistance ($R_S$ and $R_P$) are chosen to precisely tune the nTron gate current.

We design our proposed NOT gate topology following a similar approach as illustrated in Fig.2(d). When the input voltage $V_{IN}$ is low (-0.7 V, $V_{GT}$ = 0V), the JJFET exhibits high resistance diverting a large portion of $I_{Bias1}$ into the nTron gate through the series resistor ($R_S$) (Fig.2(e)). The resulting current through the nTron gate is sufficient to switch the nTron channel to the resistive state and produce a high voltage (0V) across the channel (Fig. 2(e)). In contrast, when $V_{IN}$ = 0V ($V_{GT}$ = 0.7 V), the JJFET exhibits lower resistance drawing sufficient current from $I_{Bias1}$ and reducing the gate current below the switching threshold (10.3 $\mu$A). In this condition, the nTron channel remains in the superconducting state, and the output voltage stays at -0.7V effectively realizing the logic inversion (Fig.2(f)). Figs. 2(g-i) illustrates the simulation waveform for our proposed Copy and NOT gates in response to a pulsed input voltage.

Fig.3 illustrates the design and functionality of our proposed 2-input NAND and NOR gates. The NAND gate is implemented by connecting two JJFETs in series under a common bias current ($I_{Bias1}$) (Fig. 3(a)). When both the inputs are high ($V_{IN1}$ = 0 V, $V_{IN2}$ = 0 V), the series connected JJFETs exhibit a low resistive path allowing a significant portion of $I_{Bias1}$ flows through them (Fig.3(c)). As a result, a small current flows into the nTron gate through the series resistance, $R$, which is insufficient to trigger a resistive transition. The nTron channel remains superconducting, and the output stays at -0.7V, corresponding to logic '0' (Fig.3(c)). For all other input combinations, at least one JJFET remains in a high-resistance state, diverting the majority of $I_{Bias1}$ into the nTron gate. This current exceeds the switching threshold (10.3 $\mu A$), driving the nTron into the resistive state and producing a high $V_{OUT}$ (0 V) representing a logic '1' at the output (Fig. 3(b)). Similarly, a 2-input NOR gate are implemented by connecting two JJFETs in parallel as illustrated in Fig. 3(d). Here, when either of the inputs is high, a low-resistive path is created through at least one JJFET. As a result, only a small portion of $I_{Bias1}$ flows into the nTron gate which is insufficient to switch the nTron channel to the resistive state keeping $V_{OUT}$ at -0.7V realizing logic '0' (Fig. 3(f)). Conversely, when both the inputs are low ($V_{IN1}$ = -0.7 V, $V_{IN2}$ = 0.7 V), the parallel JJFETs present high resistance, diverting major portion of $I_{Bias1}$ towards the nTron gate. This drives the nTron channel to a resistive state resulting in an output voltage of 0V, representing logic '0' at the output (Fig. 3(e)). Figs. 3(g-j) illustrates the simulation waveform for our proposed NAND and NOR gates in response to pulsed input voltages at the two inputs.

Building upon the successful implementation of fundamental logic gates such as NAND and NOR, we now extend our design framework to realize a 3-input majority gate—a key element in both classical and quantum-reversible logic architectures[27]. The majority gate outputs a logical '1' when at least two of its three inputs are high (Fig. 4(b)) [28], making it versatile and universal logic primitive. This gate is particularly significant in quantum fault-tolerant systems where majority logic plays a central role in constructing error-resilient quantum circuit [24]. A unique advantage of the majority gate is its reconfigurability: by selecting specific input combinations or fixing certain inputs, it can emulate other logic gates such as AND, OR, or NAND. This functional flexibility makes it a powerful building block for logic synthesis and optimization (Fig. 4(b,c)). The schematic of our proposed majority gate is shown in Fig. 4(a). The design consists of three JJFETs connected in series, sharing a common bias current $I_{Bias1}$. The bias-current is carefully chosen such that when two or more JJFETs are in a low-resistance state, the total current flowing into the nTron gate remains below its switching threshold (10.3 $\mu A$). Under this condition, the nTron channel stays superconducting, resulting in a low output voltage (-0.7 V) (Figs.4(d-g)). To implement the majority function, we cascade an inverter stage at the output. This inverter flips the low voltage (-0.7 V) to a high voltage (0V) representing a logic '1' at the final output. Conversely, when fewer than two inputs are high (i.e., zero or one), the gate current exceeds the nTron's threshold, switching it to the resistive state and causing the cascaded nTrons to generate a low output—again resulting in logic '0' after inversion (Figs.4(d-g)). By appropriately configuring three JJFET devices and carefully tuning the current biasing conditions, we ensure that the output reflects the majority of the input states while maintaining compatibility with the voltage-level and fan-out conventions established in our earlier designs.

Finally, to demonstrate the fanout capability of our proposed JJFET-based logic topologies, we implement a **2-input XOR gate** using the previously designed **NAND, NOR, and NOT gates**, as shown in Fig. 5(a). The corresponding simulation waveforms are presented in Figs. 5(b–e). The results clearly confirm that the proposed logic gates can be successfully cascaded to form multistage Boolean logic circuits, validating their suitability for complex cryogenic logic design.

## Discussion and Conclusion

In this work, we have proposed and demonstrated a family of voltage-controlled superconducting logic circuits based on quantum-enhanced JJFETs integrated with nTron. This framework directly addresses two longstanding limitations of conventional current-controlled superconducting logic: the lack of fanout capability and the absence of voltage gain. By leveraging the strong gate dependence of critical current in a novel JJFET architecture—realized using an InAs/GaSb heterostructure that supports an excitonic insulator phase transition—we achieve enhanced gate tunability, laying the foundation for transistor-like behavior in superconducting logic.

Our developed Verilog-A-compatible model accurately captures the nonlinear, gate-dependent electrical characteristics of the JJFET enabling accurate simulation of device behavior and its integration into circuit-level designs. To resolve the inconsistency of the input and output voltage levels, we incorporated nTron-effectively restoring logic levels and achieving cascadability by amplifying mV signals to voltage swings of 0.7 V. Through circuit-level simulations, we validated the functionality of fundamental Boolean gates including NOT, NAND, NOR, and a 3-input majority gate. We further demonstrated the cascadability of our gate library by constructing a 2-input XOR gate using the proposed NAND, NOR, and NOT gates, confirming complex multistage logic building capability. The proposed hybrid architecture offers a viable path toward scalable, energy-efficient, and voltage-controlled superconducting logic beyond flux-based paradigms and provides a platform for integrating superconducting logic with quantum processors, low-power AI accelerators, and other cryogenic computing applications. Future work will focus on experimental realization of the proposed logic topologies, optimization of the JJFET-nTron interface for lower energy-delay product, and architectural exploration of large-scale logic systems built from these devices.

**Data Availability**

The data that supports the plots within this paper and other finding of this study are available from the corresponding author upon reasonable request.

**Author Contributions**

M.M.I. conceived the idea, designed the logic gates, and performed the simulations. S.A, and A.A. analyzed and helped finalizing the designs. J. P. M., D. M., W. P. provided insights into the device's characteristics. A.A. supervised the project. All authors commented on the results and wrote the manuscript.

**Competing Interests**

The authors declare no competing interests.

**Acknowledgement:**
The work at Sandia National Laboratories (SNL) is supported by an LDRD project. DF is partially supported by the SMART program at SNL. SNL is a multi-mission laboratory managed and operated by National Technology & Engineering Solutions of Sandia, LLC (NTESS), a wholly owned subsidiary of Honeywell International Inc., for the U.S. Department of Energy's National Nuclear Security Administration (DOE/NNSA) under contract DE-NA0003525. This written work is authored by an employee of NTESS. The employee, not NTESS, owns the right, title and interest in and to the written work and is responsible for its contents. Any subjective views or opinions that might be expressed in the written work do not necessarily represent the views of the U.S. Government. The publisher acknowledges that the U.S. Government retains a non-exclusive, paid-up, irrevocable, world-wide license to publish or reproduce the published form of this written work or allow others to do so, for U.S. Government purposes. The DOE will provide public access to results of federally sponsored research in accordance with the DOE Public Access Plan.

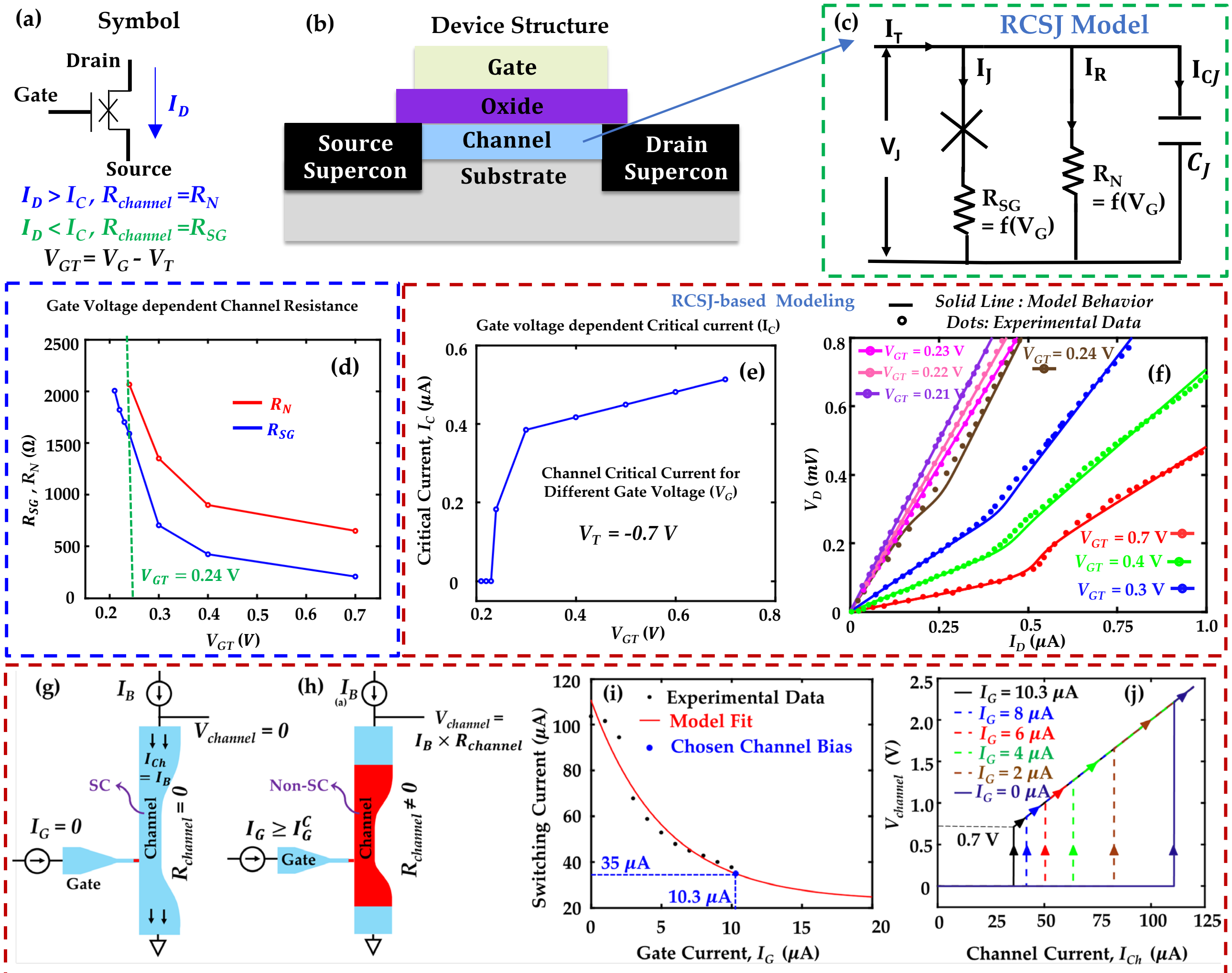

**Fig. 1. (a)** Symbol and **(b)** device structure of the Josephson Junction Field-Effect Transistor (JJFET), where superconducting Tantalum contacts define the source and drain, and the channel consists of a zero-gap InAs/GaSb heterostructure. **(c)** Equivalent RCSJ model for the quantum-enhanced JJFET. **(d)** Gate voltage ($V_G$) dependence of channel resistance, showing sub-gap resistance ($R_{SG}$) and normal resistance ($R_N$) across the critical current ($I_C$). Here, $V_{GT} = V_G - V_T$. **(e)** Gate voltage ($V_G$) dependent modulation of critical current ($I_C$). **(f)** Simulated drain voltage ($V_D$) vs drain current ($I_D$) characteristics for different $V_{GT}$. The compact model characteristics (solid line) are plotted alongside with the experimental data reported in [16] (dotted data). Gate-controlled channel switching of nTron from **(g)** superconducting (SC) to **(h)** non-superconducting state. **(i)** Channel switching current for different gate current ($I_G$) [21] with model fitting. The blue dot defines our chosen point for our proposed logic design. **(j)** Channel voltage vs Channel current ($I_C$) for different $I_G$.

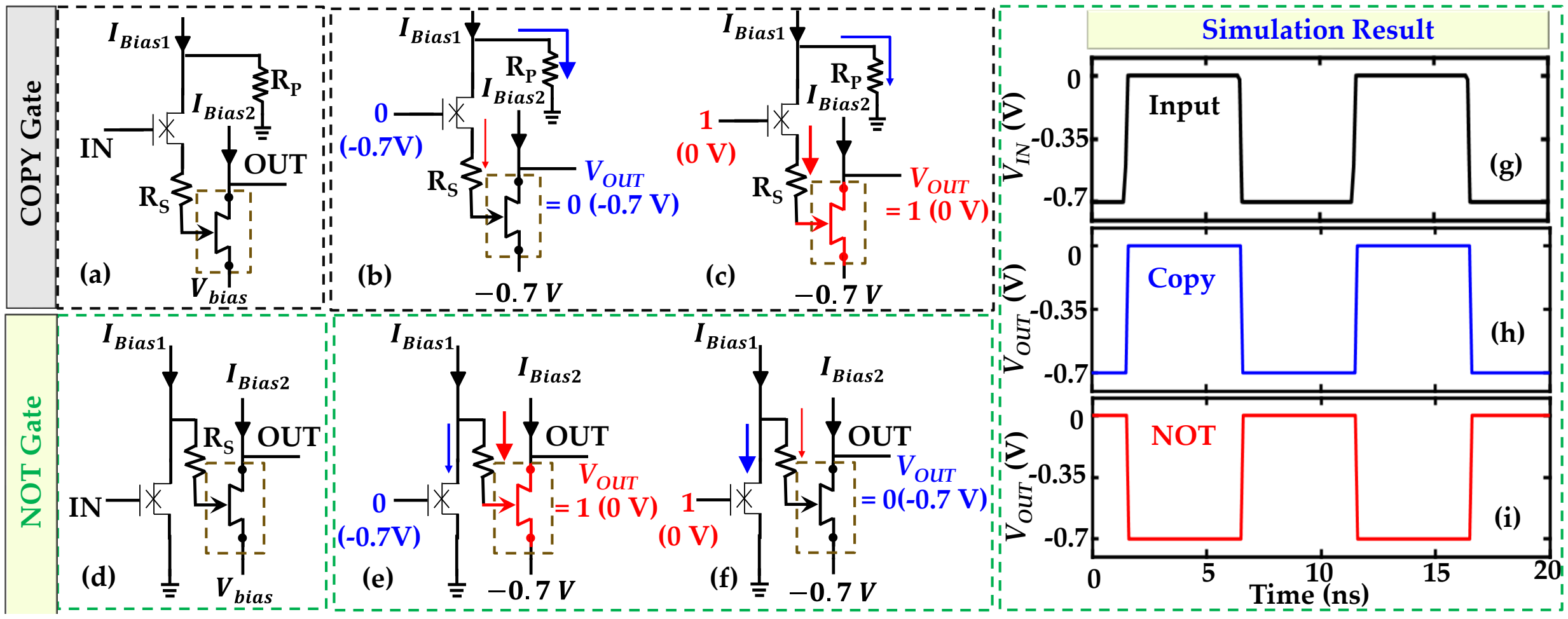

**Fig. 2. (a)** Schematics of our proposed Copy gate. Working principle of our Copy gate for **(b)** logic '0' and **(c)** logic '1'. **(d)** Schematics of our proposed NOT gate. Working principle of our proposed NOT gate for **(e)** logic '0' and **(f)** logic '1'. **(g)-(i)** Simulated results for our proposed Copy and NOT gate.

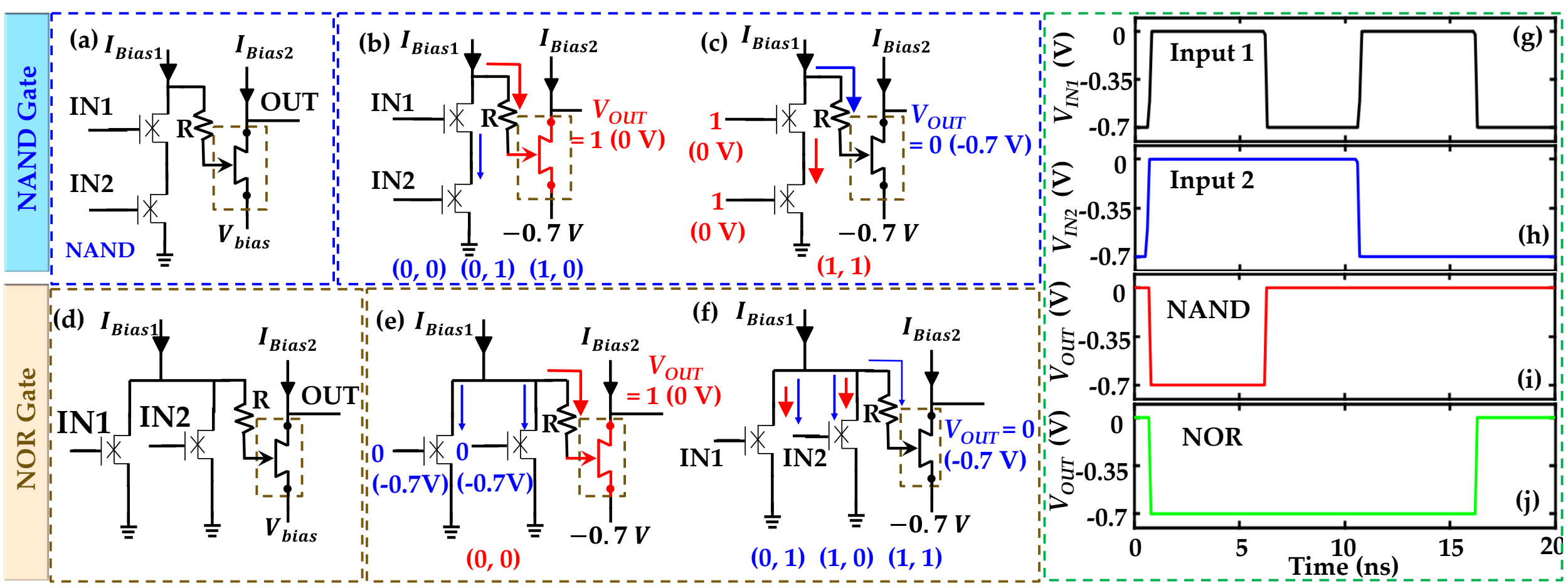

**Fig. 3. (a)** Schematics of our proposed 2-input NAND gate. Working principle of 2-input NAND gate for **(b)** logic (0, 0), (0, 1), (1, 0) and **(c)** logic (1,1) at the inputs. **(d)** Schematics of our proposed 2-input NOR gate. Working principle of 2-input NOR gate for **(b)** logic (0, 0) and **(c)** logic (0, 1), (1, 0), (1, 1) at the inputs. **(g)-(i)** Simulated results for our proposed 2-input NAND and 2-input NOR gate.

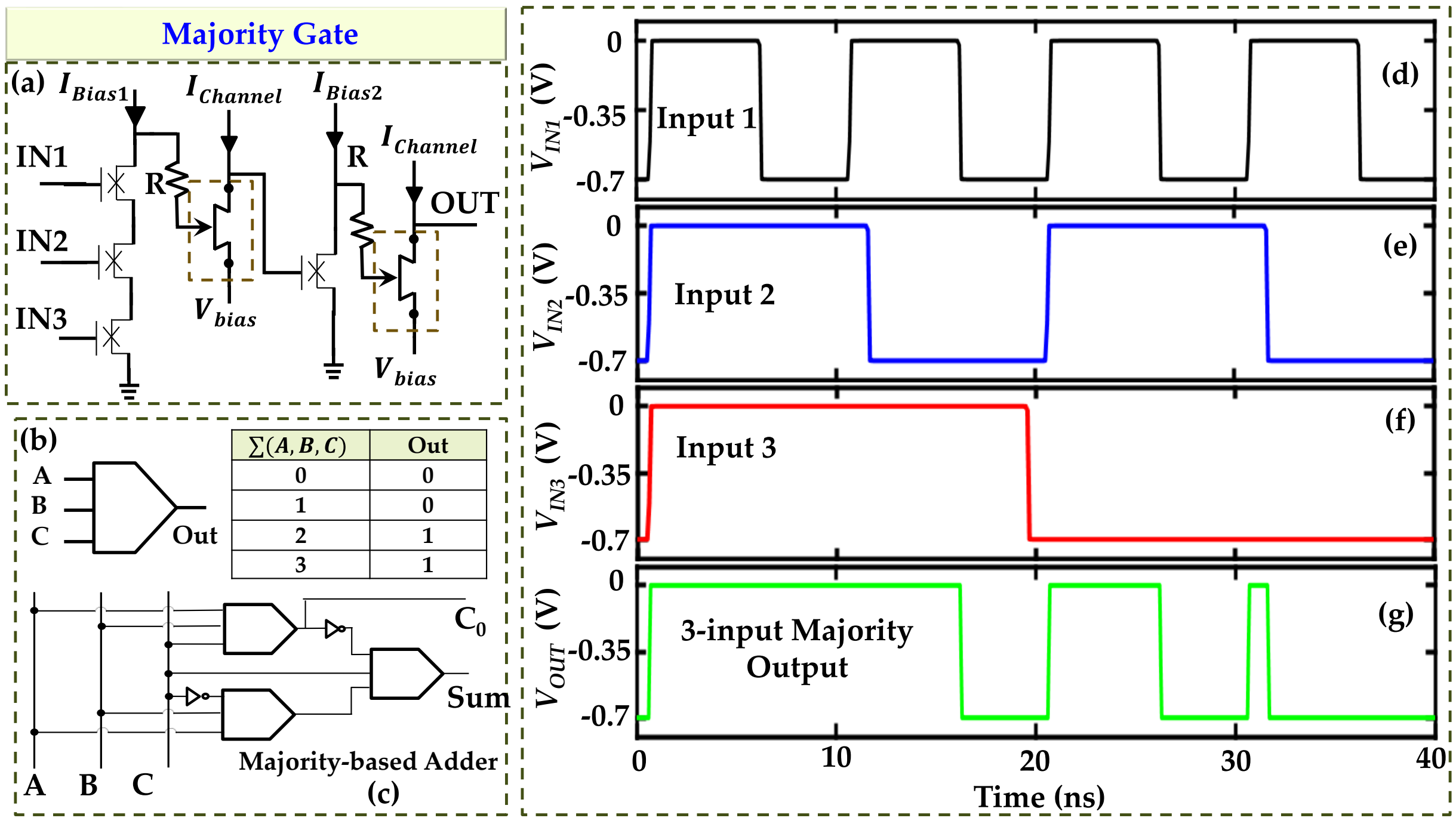

**Fig. 4. (a)** Schematics of our proposed 3-input Majority gate. **(b)** Symbol and truth table of a 3-input Majority gate. **(c)** A full adder block implemented by a 3-input majority gate. **(d)-(g)** Simulated results for our proposed 3-input Majority gate.

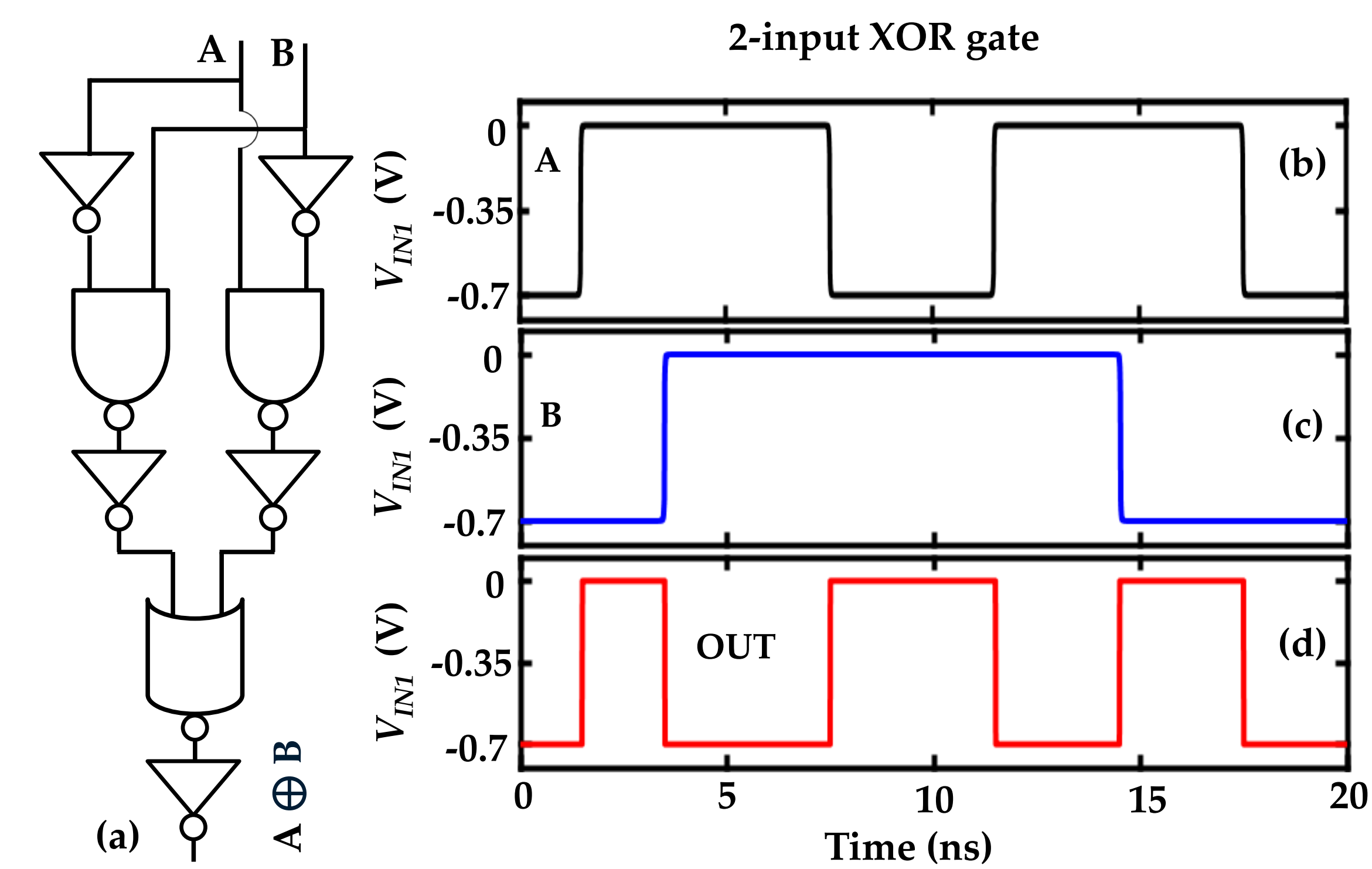

**Fig. 5. (a)** Circuit schematic of a 2-input XOR gate implemented using cascaded NOT, NAND, and NOR gates. **(b)–(d)** Simulated output waveforms demonstrating correct XOR logic behavior.